# Energy Efficient (EECP) Clustered Protocol for Heterogeneous Wireless Sensor Network

Surender Kumar
*Department of Computer Engg*
*UPES, Dehradun &university*

Manish Prateek
*Department of Computer Engg*
*UPES, Dehradun &university*

Bharat Bhushan
*Department of Computer Sc. & Appl.*
*Khalsa College, Yamunanagar*

*Abstract*— *Energy Conservation and prolonging the life of Wireless Sensor Network is one of the major issues in the wireless sensor network as sensor nodes are highly energy constrained devices. Many routing protocols have been proposed for sensor network, especially cluster based routing protocols. Cluster based routing protocols are best known for its energy efficiency, network stability and for increasing the life time of the sensor network. Low Energy Adaptive Clustering Hierarchy (LEACH) is one of the fundamental protocols in this class. In this research paper we propose a new energy efficient cluster based protocol (EECP) for single hop heterogeneous wireless sensor network to increase the life of a sensor network. Our sensor protocol use the distance of the sensor from the sink as the major issue for the selection of a cluster in the sensor network.*

## I INTRODUCTION

With the recent advances in the MEMS and wireless communication technology low cost and low power consumption tiny sensors are available. Sensors collect the data from its surrounding area, carry out simple computations, and communicate with other sensors or with the base station (BS) [1]. Main challenge for the wireless sensor network is the energy efficiency and stability because the battery capacities of the sensor nodes are limited and due to harsh deployment of the sensor nodes, it is impractical to change them. Clustering of the network is an alternative to increase the energy efficiency and performance of the network. In hierarchical architecture nodes are divided in to a number of clusters and a set of nodes are periodically elected as the cluster head (CH). Cluster heads collect the data from non-CHs nodes, aggregates the data and then send it to the base station [2] [3] .Clustering thus evenly distribute the energy load of the network, reduce the energy consumption and there upon prolonging the lifetime of the network [4], [5], [6], [7], [8], [9], [10], [12].

In this paper, we propose and evaluate a novel energy efficient cluster based protocol (EECP) for heterogeneous network. In this approach a new Cluster head (CH) election mechanism is suggested based on the initial energy of nodes and the distance between them and the sink. This protocol is an improved version of LEACH protocol presented in [4] and simulation result shows that our scheme is more efficient than LEACH.

The rest of this paper is organized as follows. Section 2 presents the related works. Section 3 describes the System Mode Used, Section 4 describes Radio Energy Model used, Section 5 describes EECP protocol, Section 6 explores the simulation results and finally paper is concluded in Section 7.

## II RELATED WORK

Routing in a wireless sensor network is more challenging than the traditional network due to sensors smaller memory, less processing power and constrained energy supply. In the past few years many new routing protocols have been devised for wireless sensor network [2], [3]. LEACH [4] is perhaps the first clustering protocol for wireless sensor network and which out performs the classical routing protocol by using adaptive clustering scheme. In this sensor node communicates with each other by using a single-hop communication. It operates in rounds and each round is further sub divided into two phases: setup phase and steady-state phase. In the setup phase each node generates a random number between 0 and 1 and if this random number is less than a particular threshold value, then the node becomes a cluster head for the current round. After the CH election, non CH nodes select a CH which is nearest to them and then CH creates a TDMA schedule for its members so that they can transmit their data to the CH. Cluster head aggregates the data received from its members and after that transmit it to the base station for further processing. In [4] an enhancement over the LEACH protocol, LEACH-centralized (LEACH-C) is also proposed. LEACH-C uses a centralized clustering algorithm, to distribute the clusters throughout the sensor field. HEED [6] is another popular energy efficient clustering algorithm which periodically elect cluster heads based on the hybrid of two parameters: residual energy of sensor nodes and intra cluster communication cost as a function of neighbor proximity or cluster density. EEHC [5] is another significant probabilistic clustering algorithm which extends the cluster architecture to multiple hops. It is a K-hop hierarchical clustering algorithm aiming at the maximization of the network lifetime. PEGASIS [8] is a chain based protocol where nodes form a chain. Each node receives or transmits data to its close neighbor only. SEP [7] is a heterogeneous-aware routing protocol in which every node independently elects itself as a cluster head based on its initial energy relative to other nodes. DEEC [9] is an algorithm in which cluster head is selected on the basis of probability ratio of residual energy and average energy of the network. In [10] strengths and weakness of many existing and new protocols are analyzed. In [12] a distance based protocol is proposed for heterogeneous network.

### III  WIRELESS SENSOR NETWORK MODEL

In this section, we describe the network model used for this research. We have assumed the following assumptions for the network model and sensors:

(i) N sensors are uniformly dispersed within a M x M square meters region (Figure 1)
(ii) All sensor nodes and base station remain stationary after deployment
(iii) Some fraction of total nodes are equipped with more energy than the normal nodes
(iv) Sensors cannot be recharged .i.e. they will die after energy is exhausted
(v) CHs performs data aggregation
(vi) Base station is not limited in energy as comparison with the energy of other nodes of network

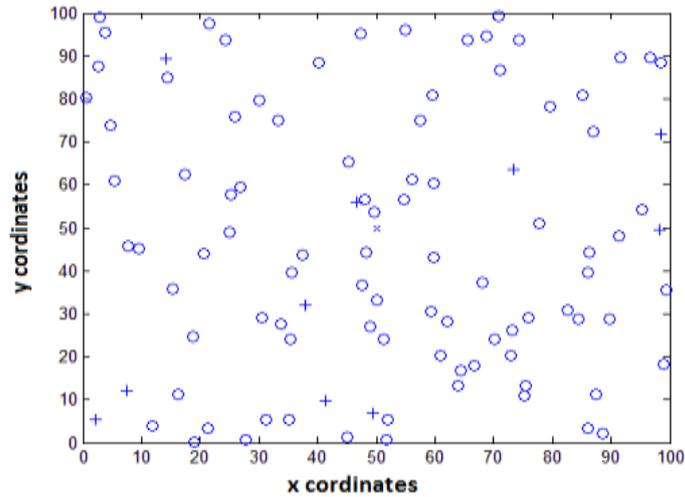

Figure.1. 100 nodes randomly deployed in the network
(o normal node, + advanced node, x Base Station)

### IV  RADIO ENERGY MODEL

Radio energy model described in [4] is used for this paper (Figure 2). Both free space ($d^2$ power loss) and the multipath fading ($d^4$ power loss) channel model is used in this model depending upon the distance between the transmitter and receiver. If the distance is less than a particular threshold value then free space model are used otherwise multipath loss model is used. The amount of energy required to transmit L bit packet over a distance, d is given by Equation (1).

$$L_{TX(L,d)} = \begin{cases} L*E_{elec} + L*\varepsilon_{fs}*d^2 & if\ (d < d_0) \\ L*E_{elec} + L*\varepsilon_{mp}*d^4 & if\ (d \geq d_0) \end{cases} \quad (1)$$

$E_{elec}$ is the electricity dissipated to run the transmitter or receiver circuitry. The parameters $\varepsilon_{mp}$ and $\varepsilon_{fs}$ is the amount of energy dissipated per bit in the radio frequency amplifier according to the distance $d_0$ which is given by the Equation (2).

$$d_o = \sqrt{\frac{\varepsilon_{fs}}{\varepsilon_{mp}}} \quad (2)$$

For receiving an L bit message the radio expends the energy given by Equation (3)

$$E_{rx}(L) = L*E_{elec} \quad (3)$$

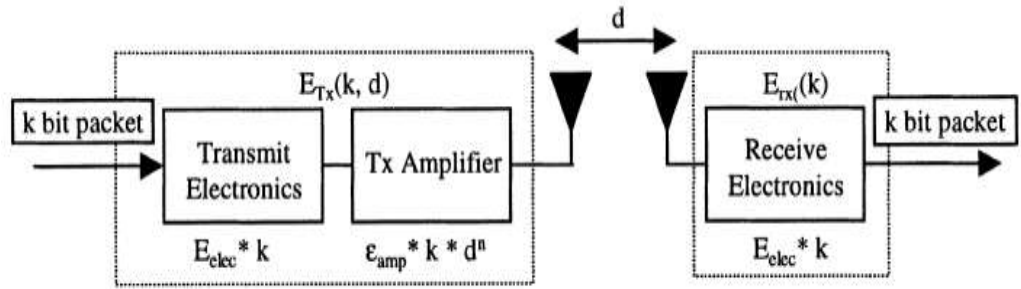

Figure 2 Radio Energy Dissipation Model

## V  EECP Protocol

In this section we describe EECP protocol which is an extension of the LEACH protocol for heterogeneous network. For taking the advantage of heterogeneity we use some sensors of higher energy than the normal nodes which we call GN (Gateway Node). Gateway nodes change the method of sending the data to the base station .The main goal of this protocol is to efficiently maintain the energy consumption of sensor network and increase the lifetime of the network. The major differences in our scheme and LEACH are as follows: (i) EECP use a new distance based probability scheme for the election of cluster head. (ii) A few percentages of the nodes in EECP have more energy than the others.The operation of EECP is divided into rounds and each round further consists of two phases

(i) Setup Phase
(ii) Steady State Phase

In the setup phase EECP elects Cluster head, form cluster and determines cluster communication schedule. For the election of cluster head sensor nodes choose a random number between 0 and 1. If the number is less than a particular threshold value then the node becomes a cluster head for the current round. If the elected cluster head is a gateway node then in Steady State phase data collected from non CHs is transmitted to the base station directly. If the elected cluster head is not a gateway node then it will calculate its distance from the base station and also from all the gateway nodes which are not elected as the cluster head in this round. If the distance of the elected cluster head from any of the gateway node (which is not selected as a cluster head in this round) is less than its distance from the sink then the elected cluster head after collecting the data from its members, aggregates the data and then send it to the nearest gateway node which further sends it to the base station.

EECP also further modified the cluster head election probability of the network. According to the radio energy dissipation model [4], the minimum amplifier energy is proportional to the square of the distance from the transmitter to the receiver. The transmission energy consumption is increased as the transmission distance increases. Thus cluster heads which are far from the base station require more energy to send the data to the BS. There is a considerably difference between the energy consumption of the nodes which are closer to the base station then the nodes which are far from the base station after the network operates for some round [12]. To get rid of this problem, we have proposed a new probability scheme so that the nodes which are far from the base station have the less chance to become cluster head in a round. After deployment base station broadcasts a "Hello Message" to all the sensor nodes at a certain power level and based on this signal strength each node computes its approximate distance ($D_i$) from the BS.

Let $D_i$ is the distance between node $S_i$ and the base station. $D_{avg}$ is the average distance between nodes and the sink. The average distance $D_{avg}$ of the nodes can be calculated by using the Equation (4)

$$D_{avg} = \frac{1}{n}\sum_{i}^{n} D_i \qquad (4)$$

The Value of $D_{avg}$ can be approximated as [12]

$$D_{avg} \simeq d_{TO\ CH} + d_{TOBS} \qquad (5)$$

Where $d_{TOCH}$ is the average distance between the node and the associated cluster head
$d_{TOBS}$ is the average distance between the cluster head and the sink. As the nodes are uniformly distributed and the sink is located in the center of the field then according to [4] [5]

$$d_{TOCH} = \frac{M}{\sqrt{2k\pi}} \quad \text{and} \quad d_{TOBS} = 0.765\frac{M}{2} \qquad (6)$$

Where *k* is the number of clusters and it can calculate by using Equation (7)

$$k = \frac{\sqrt{E_{fs}}}{\sqrt{E\,mp}} \frac{\sqrt{N}}{\sqrt{2\Pi}} \frac{M}{d^2_{BS}} \qquad (7)$$

From Equations (5), (6), (7) we can calculate the value of $d_{TOCH}$ $d_{TOBS}$ and finally $D_{avg}$

Now if the distance $D_i < D_{avg}$ then EECP use the Equation 8 and if the distance $D_i >= D_{avg}$ then Equation 9 is used for calculating the threshold value T(n). EECP use the Equation 9 only for the calculating the threshold value T(n) for the advanced (gateway) nodes.

$$T(n) = \begin{cases} \dfrac{P}{1 - P \times (r \bmod \frac{1}{P})} \times (1 - \dfrac{D_i}{D_{avg}}) & \text{if } n \varepsilon\ G \\ 0 & \text{otherwise} \end{cases} \quad (8)$$

Else

$$T(n) = \begin{cases} \dfrac{P}{1 - P \times (r \bmod \frac{1}{P})} & \text{if } n \varepsilon\ G \\ 0 & \text{otherwise} \end{cases} \quad (9)$$

## VI  Simulation Results

We have evaluated the performance of EECP with LEACH. To be fair with our evaluation we have introduced some level of heterogeneity in LEACH. For evaluation we have used 100 x 100 square meters region with 100 sensor nodes and base station is located in middle of the region as shown in Figure 1. We denote the normal nodes by using the symbol (o), advanced nodes (Gateway) with (+) and the Base Station by (x). The radio parameters used for the simulations are given in TABLE 1.The performance metrics used for evaluating the protocols are: (i) Network Lifetime: this is the time interval from the start of the operation till the last node alive (ii) Stability Period: this is the time interval from the start of the operation until the death of the first alive node  (iii) Number of Alive Nodes per round (iv) Number of Cluster Heads formed per round.

| Parameter | Value |
| --- | --- |
| $E_{elec}$ | 5 nJ/bit |
| $\varepsilon_{fs}$ | 10 pJ/bit/m$^2$ |
| $\varepsilon_{mp}$ | 0.0013 pJ/bit/m$^4$ |
| $E_0$ | 0.5 J |
| $E_{DA}$ | 5 nJ/bit/message |
| Message Size | 4000 bits |
| $P_{opt}$ | 0.1 |
| do | 70m |

Table 1    Radio Parameters used in Model

We test our protocol by introducing various parameters of heterogeneity in the system. Figure 4 shows that network lifetime of EECP is more than LEACH as first and last node dies earlier in LEACH as compared to EECP. Stability period of EECP is also more as the first node dies later in EECP. Figure 5 shows that no of alive nodes are more in EECP as compared to LEACH. After 3500 rounds no of alive nodes are more in LEACH with heterogeneity but in the end EECP overtakes LEACH and throughput of the network is also more in EECP as compared to LEACH. Figure 6 shows that number of packets send to base station per round is more in EECP as compared to LEACH. Thus throughput of the network is more for EECP as compared to LEACH.

VII    Conclusions

EECP is an extension of LEACH which takes the full advantage of heterogeneity. It improves the network lifetime, stable region and throughput of the network. For taking the full advantage of heterogeneity we have introduced some high energy nodes in the network which become gateway for transmitting the data of their cluster members to the base station. Further for increasing the energy efficiency of the network we have introduced a distance based probability scheme so that nodes which are near to the sink have the higher chances to become the cluster head. In this paper we have assumed that high energy nodes are distributed randomly. The future work includes introducing some more level of hierarchy and mobility in the network.

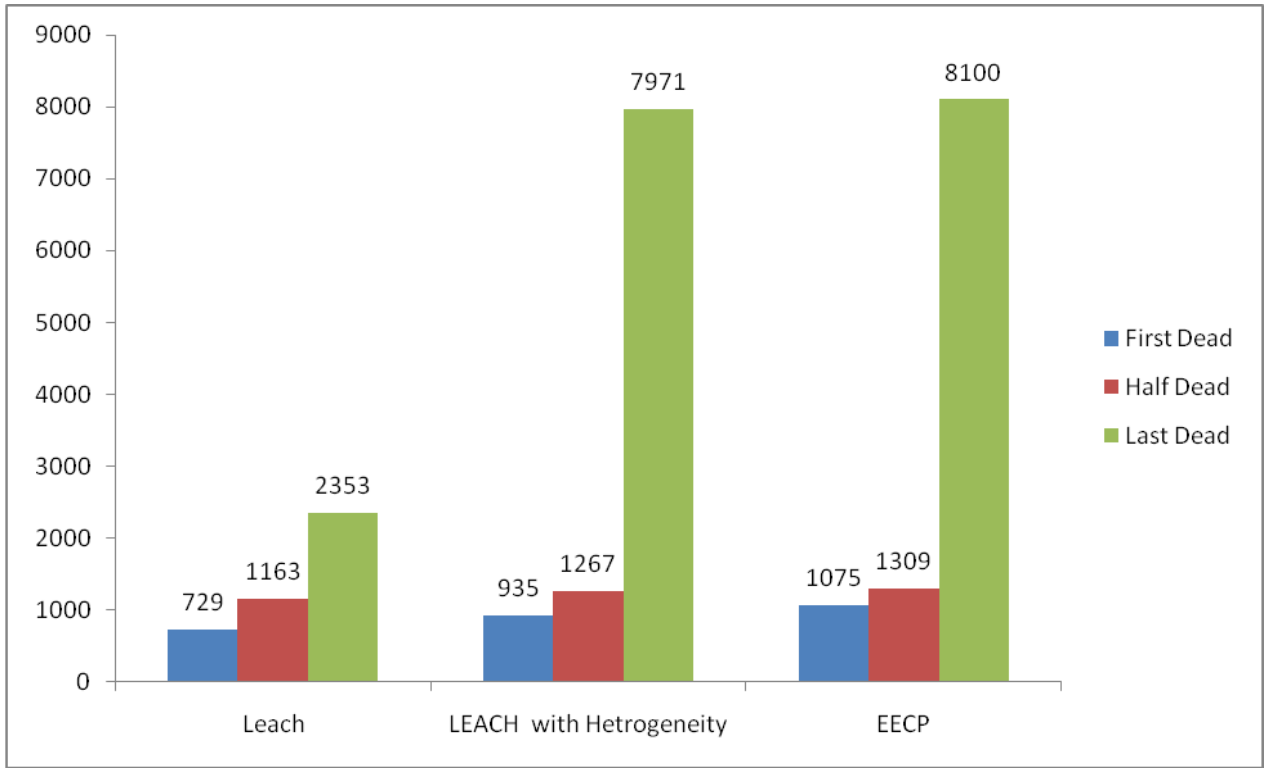

Figure 4 (Round For First,Half,Last Dead Node)

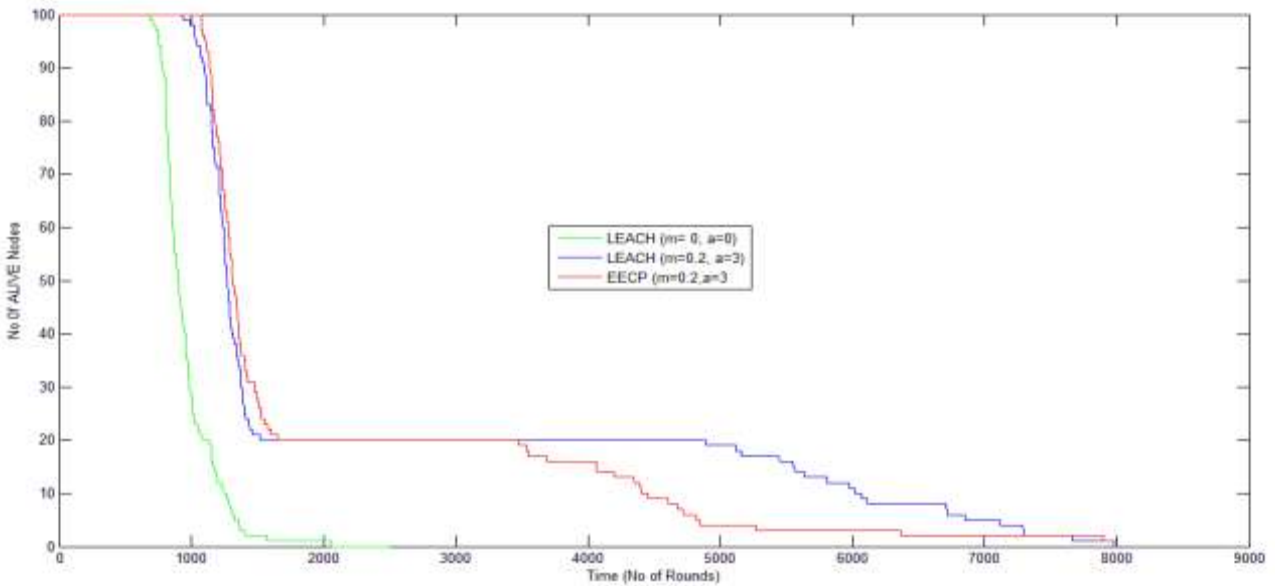

Figure 5 ( No. of Alive Nodes)

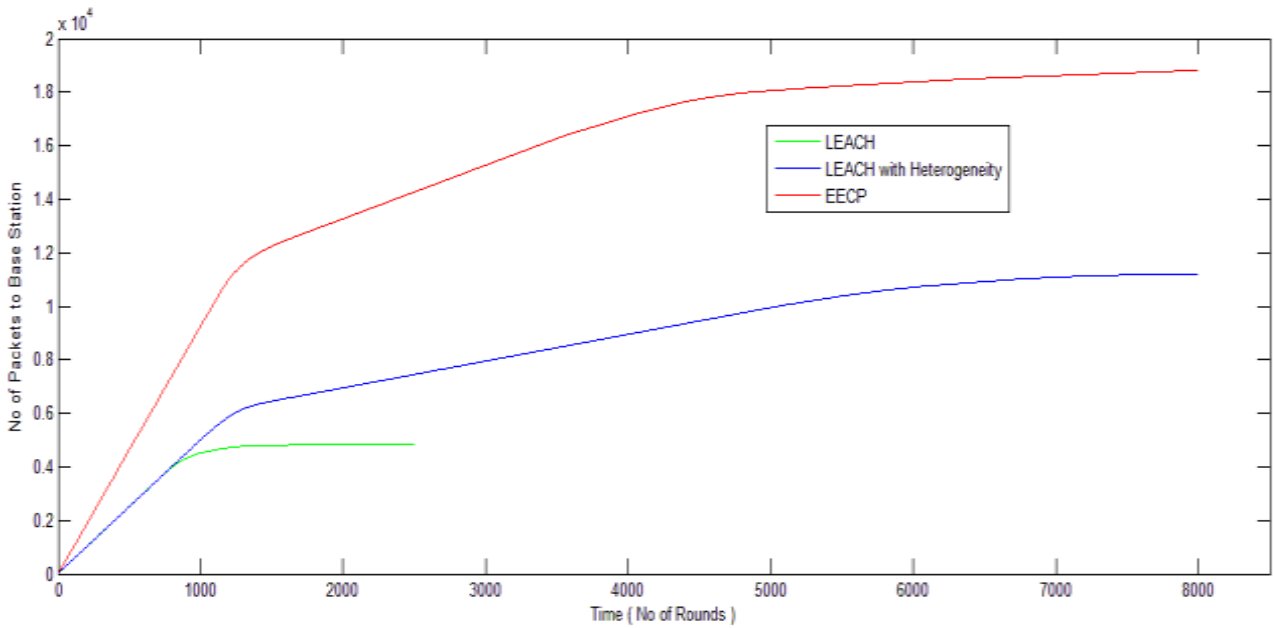

Figure 6 (No of Packets send to Base Station)

**REFERENCES**

[1]  I. F. Akyildiz, W.Su, Y. Sankarasubramaniam, and E. Cayirci, "A survey on sensor networks", IEEE Communications Magazine, vol. 40, no. 8 pp. 102–114, Aug. 2002

[2]  Kemal Akkaya, Mohamed Younis, "A Survey on routing protocols for Wireless Sensor Networks", Ad Hoc Networks 3 (2005) 325-349, 2005

[3]  S.K. Singh, M.P. Singh, and D.K. Singh, "A survey of   Energy-Efficient Hierarchical Clustered Routing in Wireless Sensor Networks", IJANA, Sept.–Oct. 2010, vol. 02, issue 02, pp. 570–580.

[4]  W. Heinzelman, A. Chandrakasan, H. Balakrishnan, "An application specific protocol architecture for Energy-efficient for wireless sensor networks", IEEE Transactions on Wireless Communications,1(4),660-670,2002

[5]  S. Bandyopadhyay, E.J. Coyle, "An Energy Efficient Hierarchical Clustering Algorithm for Wireless Sensor Networks," in: Proceeding of INFOCOM 2003, April 2003

[6]  O. Younis, S. Fahmy, "HEED: A Hybrid, Energy- Efficient, Distributed clustering approach for Ad Hoc sensor networks", IEEE Transactions on Mobile Computing 3 (4) (2004) 366–379

[7]  G. Smaragdakis, I. Matta, A. Bestavros, "SEP: A Stable Election Protocol for clustered heterogeneous wireless sensor networks", in: 2nd International Workshop on Sensor and Actor Network Protocols and Applications (SANPA 2004), 2004.

[8]  S. Lindsey, C. Raghavendra,"PEGASIS: Power- Efficient Gathering in Sensor Information Systems," IEEE Aerospace Conference Proceedings, 2002, Vol. 3. No. 9-16, pp. 1125-1130.

[9]  L. Qing, Q. Zhu, M. Wang, "Design of a distributed energy-efficient clustering algorithm for heterogeneous wireless sensor networks". ELSEVIER, Computer Communications 29, pp 2230-2237, 2006

[10]  S. Xun, "A Combinatorial Algorithmic Approach To Energy Efficient Information Collection in Wireless Sensor Networks", ACM Transactions on Sensor Networks, 3(1), 2007

[11]  Dilip Kumar, Trilok C. Aseri, R.B. Patel, "EEHC: Energy efficient heterogeneous clustered scheme for wireless sensor networks", .Elsevier, Computer Communications 32 (2009) 662–667.

[12]  Said Benkirane, Abderrahim Benihssane, M.Lahcen Hasnaoui, Mohamed Laghdir. "Distance-based Stable Election Protocol (DB-SEP) for Heterogeneous Wireless Sensor Network". IJCA, Nov. 2012